\documentclass[%
reprint,
amsmath,
amssymb,
aps,
pra,
superscriptaddress,
preprintnumbers,
prstab,
prstper]{revtex4-2}
\usepackage{graphicx}
\usepackage{dcolumn}
\usepackage{bm}

\usepackage[english]{babel}
\usepackage[utf8]{inputenc}
\usepackage{amssymb,amsfonts,amsmath,mathtools,mathrsfs}
\usepackage{relsize}
\usepackage{mdframed}
\usepackage{enumitem}
\usepackage{graphicx}
\usepackage{url,hyperref}
\usepackage[table,usenames,dvipsnames]{xcolor}
\usepackage{gensymb}
\usepackage{grffile}
\usepackage{float}

\hypersetup{
colorlinks=true,
linkcolor = blue,
filecolor = blue,
urlcolor = Blue,
citecolor = blue,
pdftitle = {antiferromagnetic antiskyrmions}
}

\usepackage[capitalize]{cleveref}
\usepackage{textcomp}
\usepackage{braket} 

\newcounter{myequation}
\makeatletter
\@addtoreset{equation}{myequation}
\makeatother

\newcounter{myfigure}
\makeatletter
\@addtoreset{figure}{myfigure}
\makeatother

\begin{document}

\title{
Engineering antiferromagnetic skyrmions and antiskyrmions at metallic interfaces}

\author{Arnob Mukherjee}
\email{ph14011@iisermohali.ac.in}
\affiliation{Department of Physical Sciences, Indian Institute of Science Education and Research (IISER) Mohali, Sector 81, S.A.S. Nagar, Manauli PO 140306, India}
\author{Deepak S. Kathyat}
\email{deepakcool6@gmail.com}
\affiliation{Harish-Chandra Research Institute, HBNI, Chhatnag Road, Jhunsi, Allahabad 211019, India}
\author{Sanjeev Kumar}
\email{sanjeev@iisermohali.ac.in}
\affiliation{Department of Physical Sciences, Indian Institute of Science Education and Research (IISER) Mohali, Sector 81, S.A.S. Nagar, Manauli PO 140306, India}

\begin{abstract}
We identify a mechanism to convert skyrmions and antiskyrmions into their antiferromagnetic (AFM) counterparts via interfacial engineering. The key idea is to combine properties of an antiferromagnet and a spin-orbit (SO) coupled metal. Utilizing hybrid Monte Carlo (HMC) simulations for a generic microscopic electronic Hamiltonian for the interfacial layers, we explicitly show the emergence of AFM skyrmions and AFM antiskyrmions. We further show that an effective spin Hamiltonian provides a simpler understanding of the results. 
We discuss the role of electronic itinerancy in determining the nature of magnetic textures, and demonstrate that the mechanism also allows for a tuning of antiskyrmion size without changing the SO coupling. 
\end{abstract}

\date{\today}
\maketitle

{\it Introduction}:
Magnetic skyrmions and antiskyrmions are topologically protected magnetization textures that have been observed
in a number of chiral magnets. Their potential use as building blocks of next-generation spintronics devices 
has motivated numerous scientific studies \cite{fert2017, tokura2017, jungwirth2016, nagaosa2013b, 
bogdanov2020, wiesendanger2016, Fert2013}. Despite many advantages associated with skyrmions, such as, enhanced
stability due to its topological protection \cite{banerjee2014,pan2020,wang2017,je2020,cortes2017, hayami2021}, ultra-low 
current density dynamics \cite{yu2012,luo2020,jonietz2010, yambe2021, hayami2019}, and large Hall current 
\cite{kurumaji2019,wang2020giant}, their transverse deflection upon application of current -- known as skyrmion
Hall effect -- presents a major bottleneck for applications \cite{chen2017skyrmion}. Therefore, reducing the 
transverse component of skyrmion velocity while retaining all of its favorable properties is an important goal
of research in this field \cite{toscano2020,Gobel2019,Akosa2019}. One of the approaches has been to tune the 
nature of the skyrmion state in such a way that the magnus force gets intrinsically canceled \cite{tretiakov2018, litzius2017skyrmion, barker2016}. These efforts have given rise to the concept of antiferromagnetic skyrmions \cite{gao2020, dohi2019, akosa2018a, rosales2015, tretiakov2019, tretiakov2021, mukherjee2020a, mukherjee2021}.

While there are now a few examples of skyrmions in insulating magnets, these textures are mostly observed in 
metals \cite{seki2012, Adams2012, kurumaji2017, bordacs2017}. Indeed, the property that skyrmions can be driven by ultra-low currents comes into play only if the host material is a metal. Therefore the challenge is not only to convert skyrmions into antiferromagnetic skyrmions, but also to do so while retaining the metallic character of the host. Recent experiments report the observation of antiferromagnetic skyrmions in metal – antiferromagnet interfaces, which can be driven efficiently by applying spin-orbit torque \cite{legrand2020}.

In this work, we explicitly show that antiferromagnetic counterparts of skyrmions and antiskyrmions can be stabilized at the interface between an antiferromagnetic insulator and a spin-orbit coupled metal. While we only discuss the case of Dresselhaus metal here, the interface engineering approach to convert skyrmions into antiferromagnetic skyrmions is applicable in general for Bloch and N\'eel skyrmions as well as antiskyrmions. 
We also discuss the role of electronic hopping in determining the nature of the topological magnetic textures. 
The results are obtained via two independent methods: (i) hybrid Monte Carlo simulations that explicitly retain the itinerancy of electrons, and (ii) classical simulations of an effective spin Hamiltonian. While the first approach provides a direct accurate treatment of electronic model, the second offers a simpler understanding of the results. 
\begin{figure}[h]
 \includegraphics[width=0.9 \columnwidth,angle=0,clip=true]{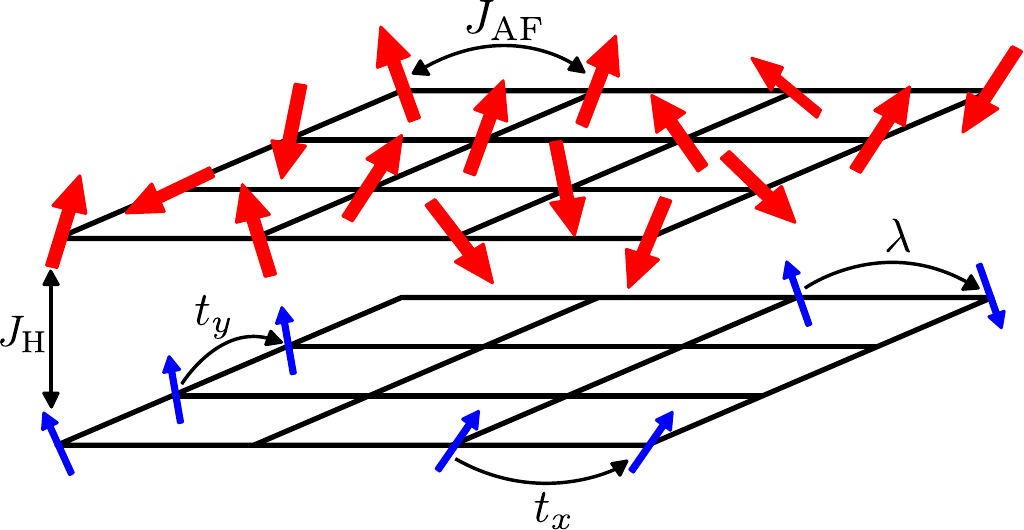}
 \caption{Schematic view of the interface between a SO coupled metal and an antiferromagnet. The upper layer consists of large magnetic moments (red arrows). The lower layer represents a SO coupled 2DEG. $J_{{\rm AF}}$ denotes the AFM coupling between localized moments, $t_{x(y)}$ denote the spin-preserving hopping along $x(y)$ directions and $\lambda$ is the strength of SO coupling visualized as spin-flip hopping. $J_{{\rm H}}$ denotes the Hund's rule coupling between the local moments and the spins of itinerant electrons.}
 \label{fig:schematic}
\end{figure}

{\it Model:}
A schematic view of the model set-up is presented in Fig. \ref{fig:schematic} where the upper layer represents the bottom surface of an antiferromagnetic insulator and the lower one the top surface of a spin-orbit coupled metal.
Note that the separation in the vertical direction is exaggerated for clarity. A prototype electronic Hamiltonian for such bilayers, with a specific choice of the type of SO coupling, is given by,
\begin{eqnarray}
	H & = & - \sum_{\langle ij \rangle,\sigma} t_{\gamma} (c^\dagger_{i\sigma} c^{}_{j\sigma} + {\textrm H.c.})
	+ \lambda \sum_{i} [\textrm{i} (c^{\dagger}_{i\downarrow} c^{}_{i+x\uparrow} + c^{\dagger}_{i\uparrow} c^{}_{i+x\downarrow}) \nonumber \\
	& & +  (c^{\dagger}_{i \downarrow} c^{}_{i+y\uparrow} - c^{\dagger}_{i\uparrow} c^{}_{i+y\downarrow})+ {\textrm H.c.}] - J_{\text{H}} \sum_{i} {\bf S}_i \cdot {\bf s}_i \nonumber \\
	& & + J_{\text{AF}} \sum_{\langle ij \rangle} {\bf S}_i \cdot {\bf S}_j,
	\label{eq:Ham_FKLM}
\end{eqnarray}
\noindent
where, $c_{i\sigma} (c_{i\sigma}^\dagger)$ annihilates (creates) an electron at site ${i}$ with spin 
$\sigma$ in the metallic layer. The first term represents electronic kinetic energy in terms of nearest neighbor (nn) hopping $t_{\gamma}$ from site $i$ to site $j = i + \gamma$ with $\gamma = x, y$, and $|t_{\gamma}|=1$ sets the basic energy scale. The second term is Dresselhaus SO coupling of 
strength $\lambda$. ${\bf s}_i = (1/2) \sum_{\sigma, \sigma^{\prime}} c_{i\sigma}^\dagger \boldsymbol{\sigma}_{\sigma \sigma^{\prime}} c_{i\sigma^{\prime}}$ is the electronic spin operator at site $i$, where, $\boldsymbol{\sigma}=(\sigma^x, \sigma^y, \sigma^z)$ is the vector of Pauli matrices. The last two terms represent the inter-layer ferromagnetic and intra-layer antiferromagnetic couplings (see Fig. \ref{fig:schematic}). ${\bf S}_i$, 
with $|{\bf S}_i| = 1$, denotes the localized spin at that site in the insulating layer. 

Note that the set-up proposed above is experimentally realizable. In fact, similar interface engineering 
ideas have been used to study topological Hall effect in chiral magnetic materials \cite{shao2019, 
zhang2016, ahmed2019, cheng2019, zhang2019}. A similar theoretical proposal to realize AFM skyrmion crystals was recently put forward \cite{gobel2017}. However, the proposed model was based on purely classical spins and hence relevant for insulators. In contrast, the present study retains the role of itinerant electrons and the mechanism is relevant to SO coupled metals. 

In case of a strong proximity effect, the coupling between localized spins in the magnetic layer and the 
electronic spins may be assumed large where double-exchange approximation ($J_{\text{H}} \to \infty$) 
provides the standard framework for analysis.
In the large $J_{{\rm H}}$ limit, we obtain the Dresselhaus double-exchange (DDE) Hamiltonian \cite{kathyat2021antiskyrmions},
\begin{eqnarray}
H_{\rm DDE} &=& \sum_{\langle ij \rangle, \gamma} [g^{\gamma}_{ij} d^{\dagger}_{i} d^{}_{j} + {\rm H.c.}] + 
J_{\text{AF}} \sum_{\langle ij \rangle} {\bf S}_i \cdot {\bf S}_j \nonumber \\
& &- h_z \sum_i S^z_i,
\label{eq:Ham-DDE}
\end{eqnarray}
\noindent
where, $d^{}_{i} (d^{\dagger}_{i})$  annihilates (creates) an electron at site ${i}$ with spin parallel to 
the localized spin. The Zeeman coupling of spins to an external magnetic field of strength $h_z$ has 
been included as the last term in Eq. (\ref{eq:Ham-DDE}). The projected hopping $g^{\gamma}_{ij} = t^{\gamma}_{ij} + \lambda^{\gamma}_{ij}$ depend on 
the orientations of the local moments ${\bf S}_i$ and ${\bf S}_j$ \cite{kathyat2020a, kathyat2021antiskyrmions}.



{\it Antiskyrmions and AFM antiskyrmions:--} We study the DDE Hamiltonian using the hybrid Monte Carlo (HMC) approach that combines the electronic diagonalization with the classical Monte Carlo \cite{kumar2006, mukherjee2015, kathyat2021a,SM}. All the results discussed below are obtained for $\lambda=0.4$. We know from the previous detailed studies that the qualitative aspects of results will not change if a more realistic smaller value of SO coupling is used instead \cite{kathyat2020a, kathyat2021antiskyrmions}. 
Presence of skyrmions or antiskyrmions is inferred from the local skyrmion densities \cite{chen2016},
\begin{eqnarray}
	\mathcal{T}_{i} & = & \frac{1}{8\pi} [ {\bf S}_i \cdot ({\bf S}_{i+x} \times {\bf S}_{i+y} ) + {\bf S}_i \cdot ({\bf S}_{i-x} \times {\bf S}_{i-y})].
\end{eqnarray}
\begin{figure}[t!]
 \includegraphics[width=0.9 \columnwidth,angle=0,clip=true]{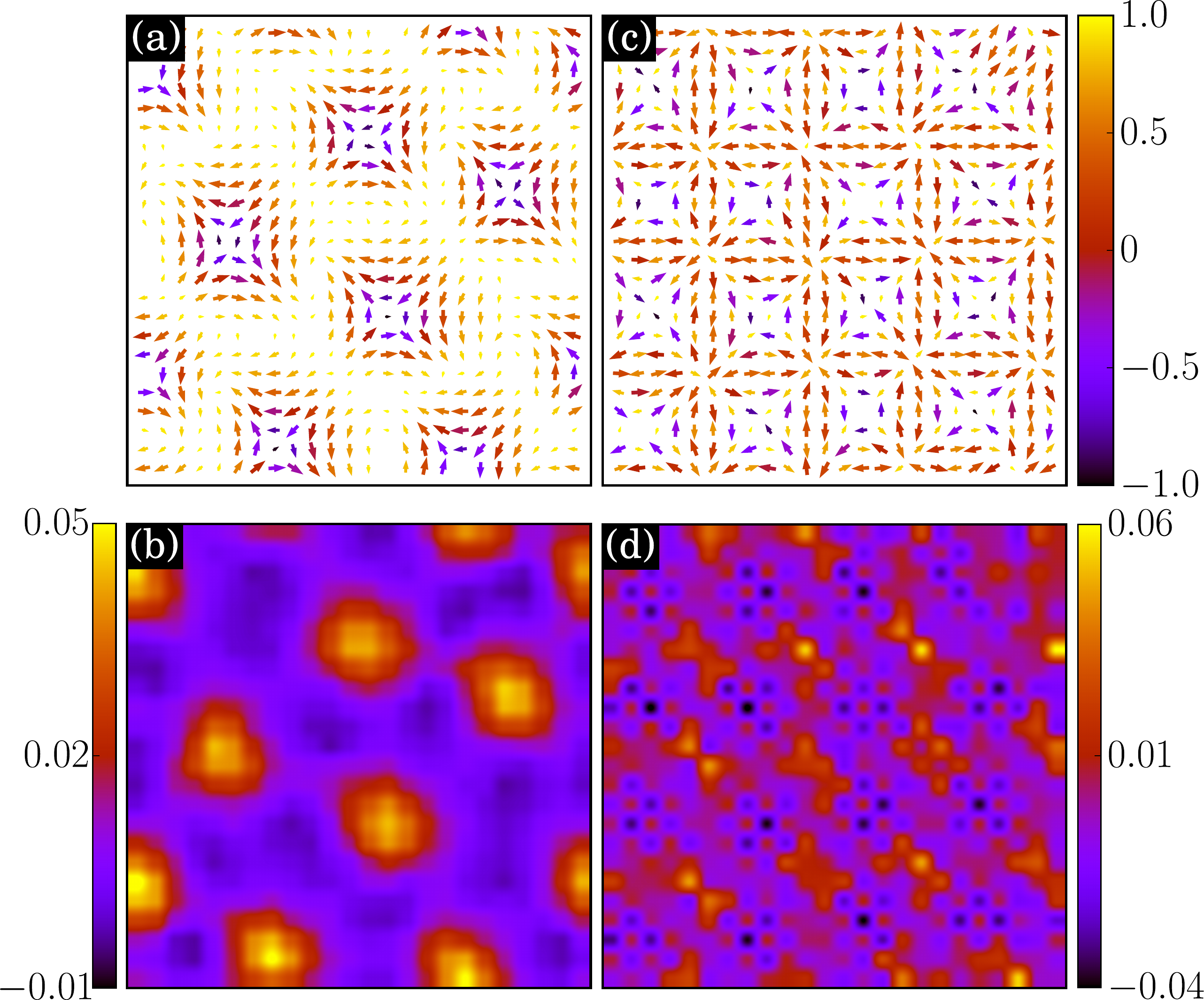}
 \caption{Snapshots of typical spin configurations, (a), (c), and the corresponding local skyrmion density, (b), (d), at $T=0.001$ for representative values of $J_{\text{AF}}$ and $h_z$ as obtained in HMC simulations: (a)-(b) $J_{\text{AF}}=0.0$, $h_z=0.03$; (c)-(d) $J_{\text{AF}}=0.2$, $h_z=0.45$. The HMC simulations are  performed  on  $24 \times 24$ lattice for an average electron filling of $0.3$ per site.}
 \label{fig:tca}
\end{figure}

Representative spin configurations in the magnetic states obtained at low temperature ($T$) via HMC simulations are shown in Fig. \ref{fig:tca}. For small $J_{\text{AF}}$, the antiskyrmion state that is present at $J_{\text{AF}} = 0$ continues to be stable \cite{kathyat2021antiskyrmions}. The formation of the antiskyrmions is confirmed by the spin configuration (see Fig. \ref{fig:tca}(a)) as well as the skyrmion density map (see Fig. \ref{fig:tca}(b)) at low-temperatures. The opposite signs of $\mathcal{T}$ and polarity on the central spin characterize the textures as antiskyrmions. We find an exotic AFM antiskyrmion crystal (AF-ASkX) state at $J_{\text{AF}}=0.2$ and $h_z=0.45$ (see Fig. \ref{fig:tca} (c)). The AF-ASkX state can be considered as a superposition of AFM and antiskyrmion configurations. This is reflected in the local skyrmion density map as a checkerboard pattern inside the antiskyrmion cores (see Fig. \ref{fig:tca} (d)). The AFM antiskyrmion phase obtained within the HMC simulations motivates a careful exploration of the phase space. Furthermore, it is also important to understand if an effective Hamiltonian for spins, obtained by integrating out the electrons, is capable of describing this conversion to AFM antiskyrmions. In the following, we present results on the effective spin-only model that allows access to much larger lattice sizes in comparison to HMC.

\begin{figure}[t!]
 \includegraphics[width=0.99 \columnwidth,angle=0,clip=true]{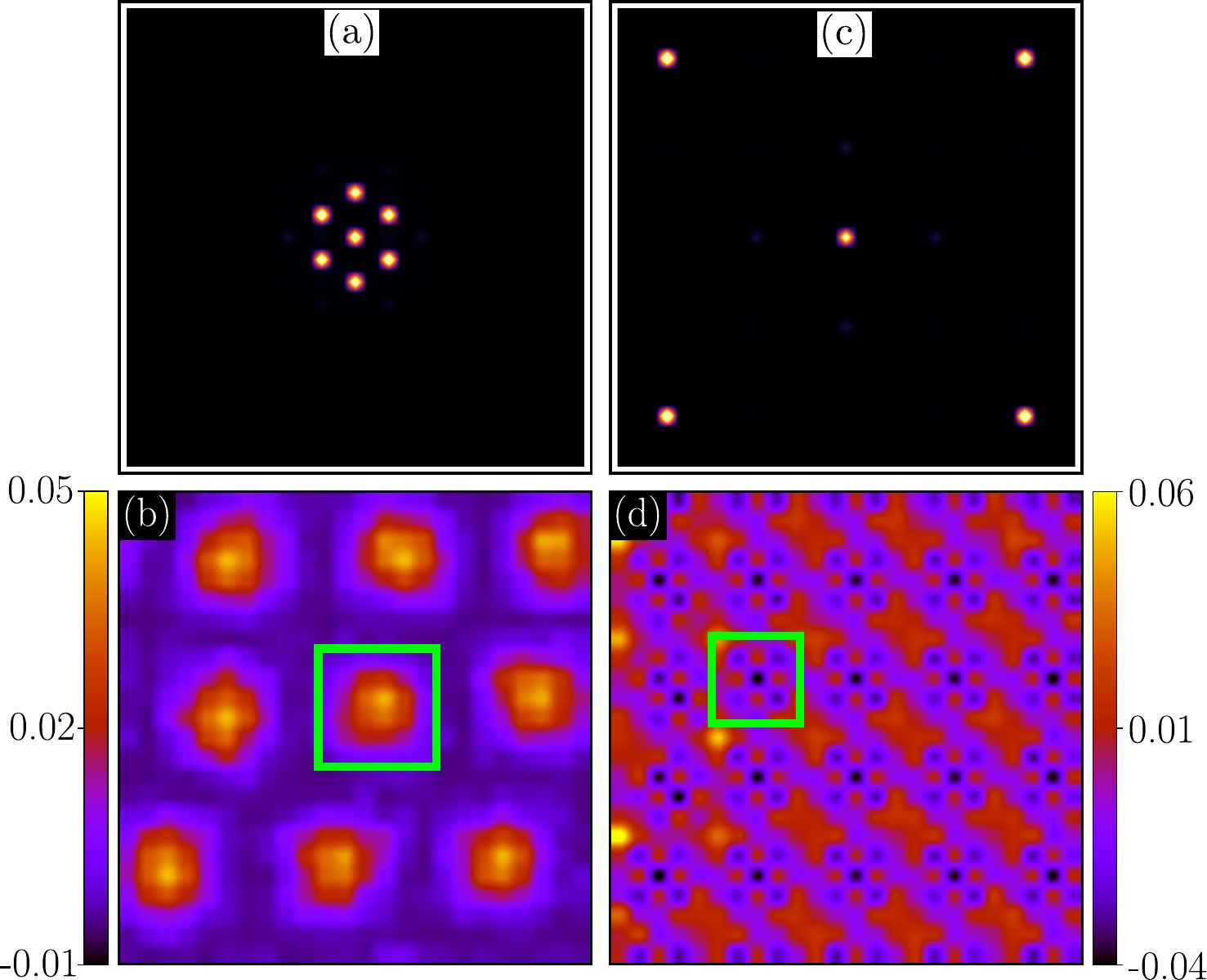}
 \caption{Locations of peak in the first Brillouin zone in the SSF for the pASk state at, (a) $J_{\text{AF}}=0.03$, $h_z=0.13$ and (c) the AF-ASkX state at $J_{\text{AF}}=0.48$, $h_z=1.1$. The lower panels (b) and (d) display the corresponding skyrmion density maps. 
 }
 \label{fig:ssf_lsd}
\end{figure}

{\it AFM antiskyrmions in the effective spin model:--}
Derivation of an effective spin model for the Hamiltonian Eq. \ref{eq:Ham-DDE} was carried out by us in a recent study \cite{kathyat2021antiskyrmions}. The effective spin Hamiltonian is given by,
\begin{eqnarray}
	H_{\rm eff} & = &-\sum_{\langle ij \rangle} D^{x(y)}_{ij} f^{x(y)}_{ij} + J_{\text{AF}} \sum_{\langle ij \rangle} {\bf S}_i \cdot {\bf S}_j  
	- h_z \sum_i S^z_i, \nonumber \\
	f^{x(y)}_{ij} & = &  1/\sqrt{2} \Big[ t^2\{1+{\bf S}_i \cdot {\bf S}_j\} -(+) 2t\lambda \hat{x} (\hat{y}) \cdot \{{\bf S}_i \times{\bf S}_j\}  \nonumber  \\
	& & + \lambda^2\big \{1-{\bf S}_i \cdot {\bf S}_j+ 2\{\hat{x} (\hat{y}) \cdot {\bf S}_i\}\{\hat{x} (\hat{y}) \cdot {\bf S}_j\} \big\} \Big]^{1/2}.
	\label{eq:ESH_DDE}
\end{eqnarray}
\noindent
While $D^{x(y)}_{ij}$ can, in principle, be $ij$ dependent, it has been shown that $D^{x(y)}_{ij} \equiv D_0 = 1$ is a very good approximation to study the magnetic phase diagrams of the model Hamiltonian Eq. \ref{eq:Ham-DDE} \cite{kathyat2020a}. 
\begin{figure}[t!]
 \includegraphics[width=0.99 \columnwidth,angle=0,clip=true]{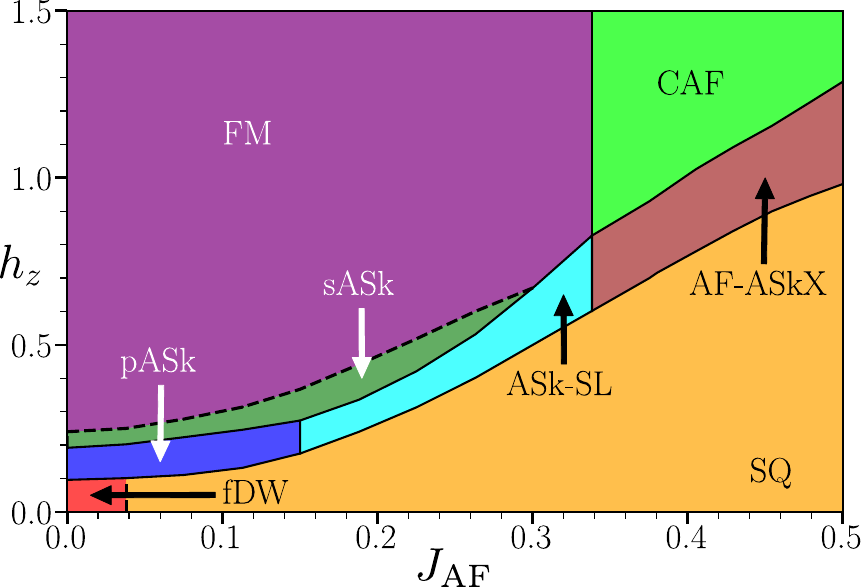}
 \caption{Low temperature ($T = 0.001$) $h_z$ vs $J_{\text{AF}}$ phase diagram. Along with saturated ferromagnetic (FM), canted antiferromagnetic (CAF), and single-Q spiral states, several exotic magnetic phases are confirmed. These are, (i) filamentary domain wall (fDW), (ii) packed-antiskyrmion (pASk), (iii) sparse-antiskyrmion (sASk), (iv) antiskyrmion square lattice (ASk-SL) and (v) AFM antiskyrmion crystal (AF-ASkX).
 }
 \label{fig:pd}
\end{figure}
We simulate $H_{\text{eff}}$ Eq. \ref{eq:ESH_DDE} using the classical Monte-Carlo approach based on the Metropolis algorithm \cite{SM}.
The $J_{\text{AF}} = 0$ limit of the model has been discussed in Ref. \cite{kathyat2021a, kathyat2020a}, where a states consisting of filamentary domain wall (fDW) structures was found to be stable at low but finite temperature and at small $\lambda$ and $h_z$.  Increasing magnetic field leads to packed-skyrmion (pSk) and sparse-skyrmion (sSk) states, and finally to the trivial saturated ferromagnetic (FM) state. The effect of AFM exchange coupling between the localized spins is summarized in the form of a low temperature phase diagram in the $h_z$-$J_{\text{AF}}$ plane (see Fig. \ref{fig:pd}). The phase boundaries are inferred from the field-dependence of magnetic susceptibility ($\chi_M$), topological susceptibility ($\chi_{\mathcal{T}}$) and spin structure factor SSF ($S_f({\bf q})$) \cite{SM}. The phase diagram consists of fDW, packed antiskyrmion (pASk), sparse antiskyrmion (sASk), antiskyrmion square lattice (ASk-SL), and an exotic AFM antiskyrmion crystal (AF-ASkX) along with the relatively trivial single-Q (SQ) spiral, saturated ferromagnetic (FM), and canted antiferromagnetic (CAF) states. 
In the small $J_{\text{AF}}$ limit the fDW state continues to be stable. Increasing $J_{\text{AF}}$ leads to a SQ spiral state which remains stable upto very large $J_{\text{AF}}$ for low magnetic fields. The instability of fDW can be understood in terms of lifting of degeneracies, by the AFM term, that stabilize the fDW state in the first place \cite{kathyat2020a}.
\begin{figure}[t!]
 \includegraphics[width=0.9 \columnwidth,angle=0,clip=true]{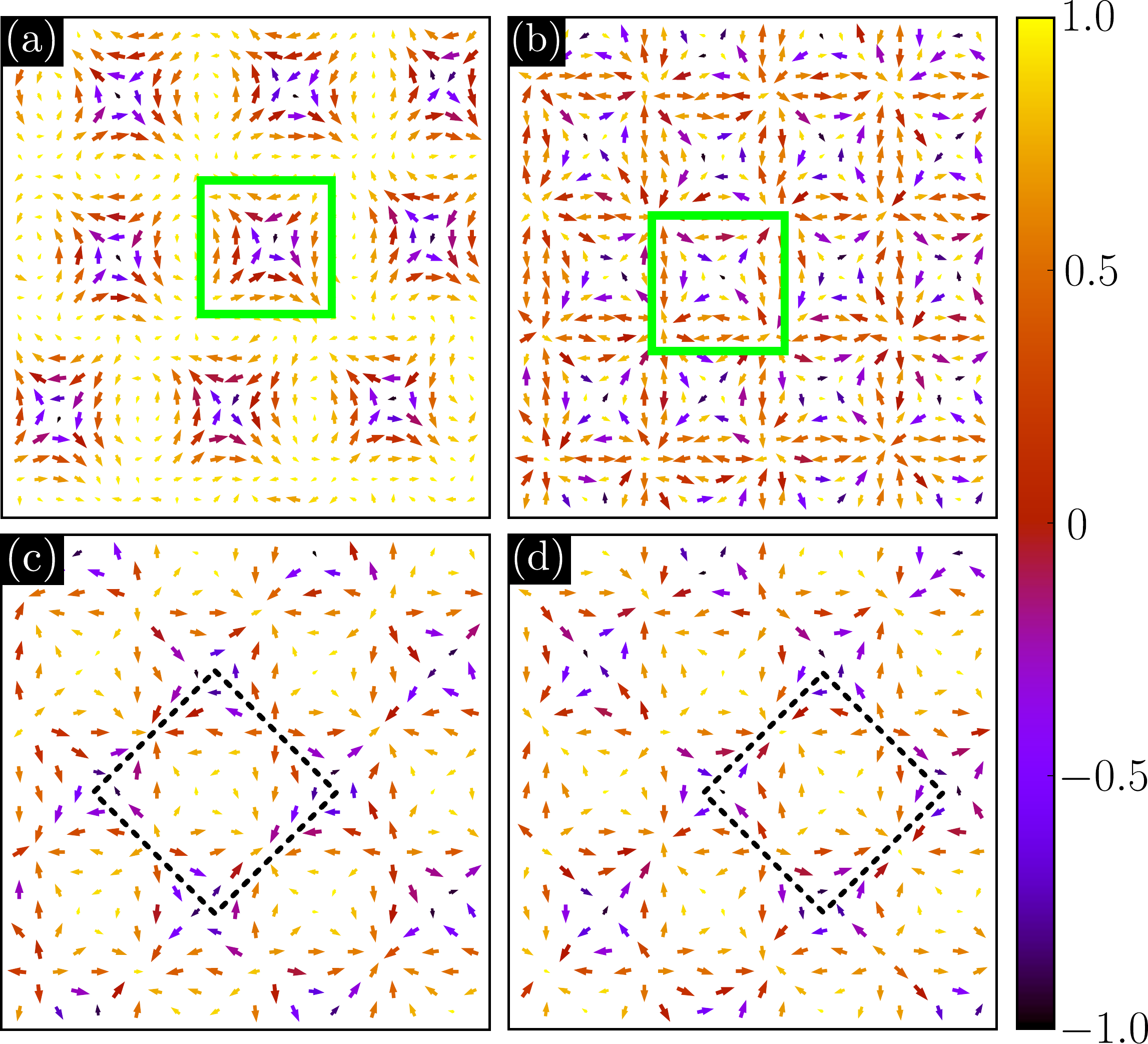}
 \caption{Snapshots of typical spin configurations at $T = 0.001$ for, (a) pASk at $J_{\text{AF}}=0.03$, $h_z=0.13$ and (b)
 AF-ASkX at $J_{\text{AF}}=0.48$, $h_z=1.1$. The planar components of the spin vectors are represented by the arrows while the z component is color coded. We have displayed a $24 \times 24$ section of the $40 \times 40$ lattice used for simulations. The green boxes enclose a single antiskyrmion or AFM antiskyrmion. The sublattice resolved spin configuration of AF-ASkX state, panel (b), are shown in (c) and (d). The black dotted squares in (c) and (d) indicate the magnetic unit cell of the tilted square lattice.
 }
 \label{fig:spin_config_eff}
\end{figure}
Increasing magnetic field leads to pASk and sASk states for small $J_{\text{AF}}$.
In the pASk state, the antiskyrmions are arranged on a triangular lattice (see Fig. \ref{fig:spin_config_eff} (a)). The hexagonal arrangement of antiskyrmions becomes more clear in the corresponding SSF plot (see Fig. \ref{fig:ssf_lsd} (a)). The hexagonal peak structure in the SSF can be understood as a superposition of three degenerate spiral states, and the peak at $(0, 0)$ corresponds to the uniform magnetization.  We also find that, within the pASk phase, increasing $J_{{\rm AF}}$ leads to a reduction in the size of antiskyrmions \cite{SM}. Generally, one needs to change the strength of SO coupling relative to the bandwidth in order to control the size of these topological magnetic textures. It is a fact of possible practical importance that $J_{{\rm AF}}$ turns out to be an independent model parameter that can alter the antiskyrmions size. For a range of values, $J_{{\rm AF}}$ seems to be playing the role of an additive factor to the SO coupling. Following our previous study, the sASk state can be identified as a metastable state at $T=0$ \cite{kathyat2021a, kathyat2021antiskyrmions}. Therefore, we mark the sASk-FM boundary with a dotted line.

Over a wide range of $J_{\text{AF}}$ values, $0.15 < J_{\text{AF}} < 0.34$, antiskyrmion square lattice (ASk-SL) state is favored in the presence of magnetic field. A similar state is known to exist in the absence of magnetic field for a narrow range of SO coupling values near $\lambda = 1.5$ and at $J_{\text{AF}} = 0$ \cite{kathyat2020a, kathyat2021a}. It is interesting to note that not only the AFM coupling increases the range of stability, it also significantly reduces the values of $\lambda$ at which the ASk-SL state exists.
At large $J_{\text{AF}}$, we find an exotic AFM antiskyrmion crystal (AF-ASkX) state in the range $0.34 < J_{\text{AF}} < 0.5$. This state can be viewed as a superposition of two spin helices, with wave vectors $(Q, Q)$ and $(-Q,Q)$ (see Fig. \ref{fig:ssf_lsd} (c)). The closeness of the peaks to $(\pi,\pi)$ reflects the AFM character of the textures. In order to better realize the underlying spin structure of the AF-ASkX, we plot the sublattice resolved spin configuration in Fig. \ref{fig:spin_config_eff} (c)-(d). The spin textures on individual sublattices in the AF-ASkX state make the tilted square arrangement of antiskyrmions apparent. The skyrmion density ($\mathcal{T}_i$) plots reveal the detailed nature of the AFM-antiskyrmion lattice states. The positive polarity of $\mathcal{T}_i$ in the pASk state is consistent with the result obtained by HMC simulation (compare Figs. \ref{fig:ssf_lsd}(b) and \ref{fig:tca}(b)). Depending on the sublattice, the central spins inside the antiskyrmions are up or down (see Figs. \ref{fig:ssf_lsd}(d) and \ref{fig:spin_config_eff} (c)-(d)) \cite{SM}.

\begin{figure}[t!]
 \includegraphics[width=0.99 \columnwidth,angle=0,clip=true]{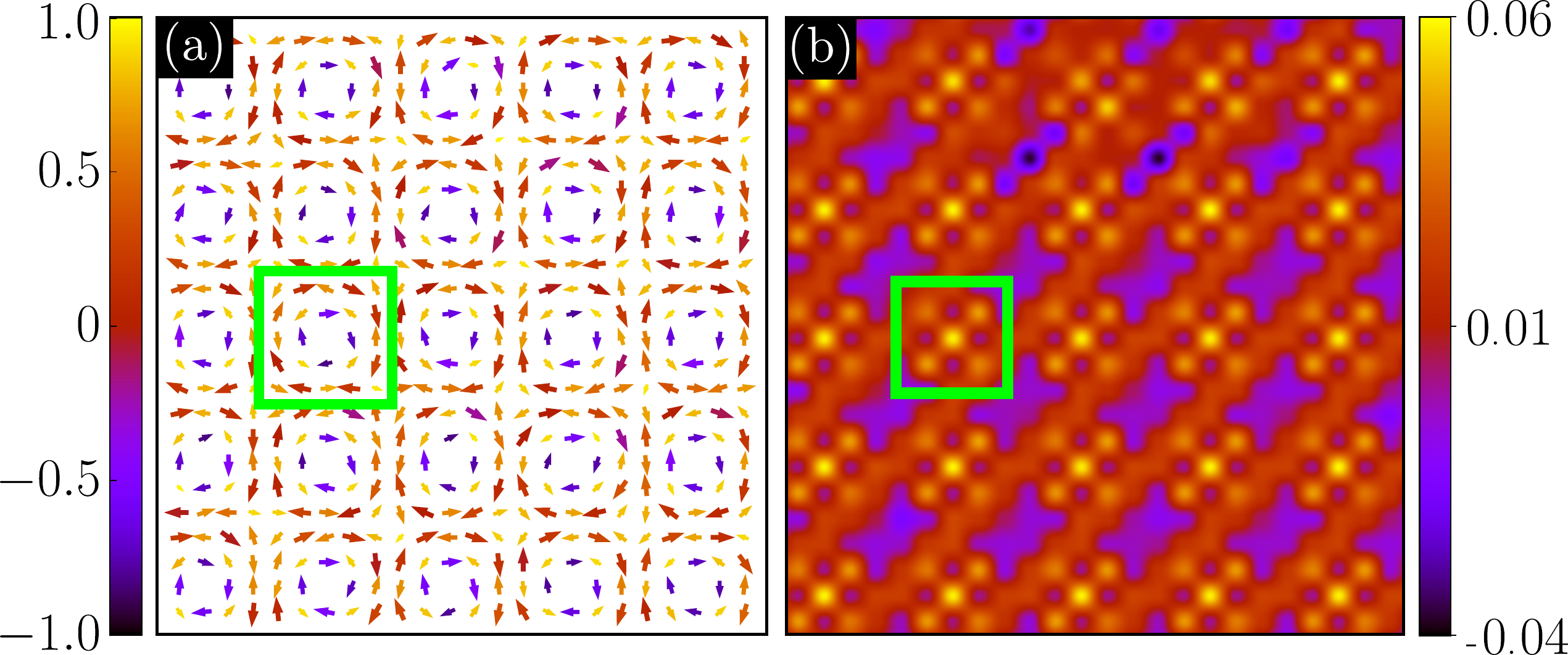}
 \caption{(a) A typical spin configuration of the AF-BSkX state. The planar spin components are represented by arrows and the information about the z component is color-coded according to the left color bar. (b) Local skyrmion density map for AF-BSkX state at $J_{\text{AF}}=0.48$, $h_z=1.1$. The results are obtained using HMC simulations of $H_{\text{DDE}}$ Eq. (\ref{eq:Ham-DDE}).
 }
 \label{fig:bloch_sk}
\end{figure}

{\it AFM antiskyrmion to AFM Bloch-skyrmion:--}
In a recent study we have shown that a relative sign change of the electronic hopping parameters leads to conversion of antiskyrmions into Bloch skyrmions \cite{kathyat2021antiskyrmions}. We explicitly show that a change of sign, $t_x \to -t_x$, in the starting Hamiltonian converts the AF-ASkX discussed above into AFM Bloch skyrmion crystal (AF-BSkX).
This is relevant for metals that show band-structure energy extremum away from the $\Gamma$ point \cite{Shanavas2016, choi2019, Nomoto2020, koretsune2015}. 
The spin configuration in the ground state of the model $H_{\text{DDE}}$ at $J_{\text{AF}} = 0.48$ and $h_z = 1.1$ is displayed Fig. \ref{fig:bloch_sk}. The Bloch character of spin textures is easily visible from the directions of the planar projections of spins.
The $\mathcal{T}_i$ maps for the AF-BSkX states (see Fig. \ref{fig:bloch_sk} (b)) display bright spots at skyrmion cores in clear contrast to dark spots seen in the case of AF-ASkX (see Fig. \ref{fig:ssf_lsd} (d)). We provide further understanding of the above conversion to AF-BSkX state with the help of a transformation on spin space of our effective spin model Eq. (\ref{eq:ESH_DDE}). The anisotropic hopping choice ($t_x \to -t_x$, $t_y \to t_y$) leads to a transformation in the spin space, $(S_x, S_y, S_z) \xmapsto{\mathcal{R}_1} (S_x, -S_y, S_z)$, such that



\noindent the energy remains invariant. It can be checked that this transformation in the spin space turns an AF-ASkX to an AF-BSkX state. This finding further establishes the importance of electronic hopping parameters, and hence the details of the band structure, in determining the intrinsic nature of topological magnetic textures.

\noindent
Although we have presented all the results for Dresselhaus SO coupling, it is easy to generalize the findings to the case of Rashba SO coupling. For $t_x = t_y$, the Rashba term prefers N\'eel skyrmions. Therefore, these will be turned into AFM N\'eel skyrmions in the presence of $J_{\text{AF}}$. For $t_x = - t_y$, the Rashba term will stabilize the antiskyrmions, and therefore  $J_{\text{AF}}$ will turn these into AFM-antiskyrmions.
Since the physics is similar, one can draw conclusions regarding the effect of $J_{\text{AF}}$ for the Rashba case even without explicit calculations. 
In Rashba case, the spin transformation matrix which maps the N\'eel skyrmions to the antiskyrmions, due to sign-change of relative hopping parameter $t_y/t_x$, is given by $(S_x, S_y, S_z) \xmapsto{\mathcal{R}_2} (-S_x, S_y, S_z)$. The transformation in the spin space which takes configurations obtained in the Dresselhaus case to those in the Rashba case can be derived from the corresponding $f_{ij}^{x(y)}$ \cite{kathyat2020a, kathyat2021antiskyrmions}. The transformation, which comes out to be $(S_x, S_y, S_z) \xmapsto{\mathcal{R}_3} (-S_y, -S_x, S_z)$, is also evident from the continuum forms of the Rashba and the Dresselhaus SO couplings. Indeed, the transformation $\mathcal{R}_3$ applied to quantum spin maps
$H_{\text{DSO}}=\lambda_D(\sigma_x k_x - \sigma_y k_y)$ on to $H_{\text{RSO}}=\lambda_R(\sigma_x k_y - \sigma_y k_x)$. Therefore, the effective Hamiltonian provides an elegant route to understand the nature of topological textures that are stable for different types of SO couplings and for different relative hopping signs.


{\it Conclusion:--}
We have proposed a general approach to tune skyrmions (antiskyrmions) into AFM skyrmions (antiskyrmions) in metallic layers with the help of interface engineering. Recent experiments report the presence of antiskyrmions in inverse Heusler metals lacking intrinsic inversion symmetry \cite{Nayak2017, Vir2019, jena2020observation}. We predict that such antiskyrmions can be turned into AFM antiskyrmions at the interface by growing an AFM insulator on top of these inverse Heusler metals. 
While the HMC simulations provide numerically accurate results about the nature of magnetic textures, the effective Hamiltonian approach provides a deeper insight into the origin of such topological textures. Furthermore, the form of effective Hamiltonian helps in identifying the mappings in spin space that solve the problem for different types of SO couplings and different choices of relative sign of $t_y/t_x$ without explicit calculations. In particular, these transformations assist us in understanding conversion of AF-ASkX into AF-BSkX which is obtained in explicit calculations. 
We also find that the ASk-SL state, which is known to be stable for rather large values of the SO coupling at $J_{\text{AF}} = 0$, is stabilized for lower values of SO coupling at finite $J_{\text{AF}}$. 
This suggests that $J_{\text{AF}}$ is acting as an additive term to the SO coupling for a range of parameters. A direct consequence, which we have explicitly verified, of this hypothesis is that a reduction in the skyrmion size should occur in systems that support an AFM coupling along with a SO coupling.
We believe that our work will motivate further experimental studies for realization of technologically superior AFM counterparts of skyrmions and antiskyrmions.

{\it Acknowledgments:--}
We acknowledge the use of computing facility at IISER Mohali.


\stepcounter{myequation}
\stepcounter{myfigure}

\renewcommand{\thefigure}{S\arabic{figure}}
\renewcommand{\theequation}{S\arabic{equation}}

\onecolumngrid

\vspace{10pt}
\hrule

\begin{center}
{\bf {\Large{Supplemental Material}}}
\end{center}

\section{Simulation Methods}

\subsection{Hybrid Monte Carlo Simulations}
In the DDE Hamiltonian (see Eq. \ref{eq:Ham-DDE} in the manuscript) the classical spin vectors are coupled with the electronic spin moments. The main difficulty in finding ground states of such coupled spin-fermion Hamiltonians is to search, amongst an exponentially large number of spin configurations, the few configurations that minimize the free energy. The Hybrid Monte Carlo simulation are well known to be best suited to handle such coupled Hamiltonians. In this approach, the classical spin variables are updated according to Metropolis algorithm, however, the electronic Hamiltonian is diagonalized at each Monte Carlo step in order to compute the energy associated with a given spin configuration. The method is computationally expansive, and hence simulations are limited to lattices $\sim 100$ sites. For simulations on larger lattices, without compromising on the accuracy, we make use of the traveling cluster approximation (TCA) \cite{kumar2006, mukherjee2015}. In this method the exact diagonalization of the fermionic Hamiltonian is performed on a smaller cluster centered around the update site, and the cluster moves along with the update site. For all the HMC results presented in this manuscript, the simulations are performed on $24 \times 24$ lattice with periodic boundary conditions using an $8 \times 8$ cluster with open boundary conditions. The `CHEEVX' subroutine of the LAPACK library is used for diagonalization of the Hamiltonian. We use $\sim 10^3$ MC steps each for equilibration and averaging at each value of temperature and Zeeman field. Other details are same as we outline below in the classical Monte Carlo simulation method.

\subsection{Classical Monte-Carlo simulations}
We study the magnetic properties of the effective spin Hamiltonian via the Classical Monte Carlo simulation technique based on conventional heat bath method \cite{binder1993monte}. Periodic boundary conditions have been employed along both the $x$ and $y$ direction. In the presence of external magnetic field, we use the zero field cooled (ZFC) protocol, where initially the temperature is lowered at zero field strength ($h_z=0$) and then at that low temperature we have increased the strength of the magnetic field. The annealing process has been achieved by reducing the temperature parameter in small steps starting at sufficiently high value to observe the phase transition from paramagnetic to ordered state. For a particular value of $T$ and $h_z$, single spin update schemes are performed by proposing a new spin configuration from a set of uniformly distributed points on the surface of a unit sphere. Note that the choice for picking a completely new orientation for spin lowers the propensity of the system to get stuck in the metastable state. The new configuration is accepted based on the standard Metropolis algorithm \cite{metropolis1953equation,hastings1970}. A Monte Carlo run at each $h_z$ and $T$ values consist of $\sim 1 \times 10^5$ Monte Carlo steps (MCSs) for equilibration and twice the number for calculating the average of the required physical observables. For detailed exploration of parameter space ($h_z$ and $J_{\text{AF}}$) we used lattice size $N = 40 \times 40$, and the stability of results is verified by simulating lattice sizes up to $N = 120 \times 120$ for some selected parameter values.

\section{Physical Observables} \label{sec:phy_ob}
The magnetic phases at low temperatures obtained in both the HMC and the Classical Monte Carlo simulations, can be visually identified by their corresponding real-space spin patterns. However, a quantitative measure becomes necessary in order to determine phase transitions between two phases. Therefore, we have calculated various physical observables to precisely identify the phase boundaries. We calculate the magnetic susceptibility ($\chi_M$) and the topological susceptibility ($\chi_\mathcal{T}$) \cite{amoroso2020}, defined as,
\begin{equation}\label{eq:obser}
\begin{split}
\chi_M           &= \frac{dM}{dh_z}, \\
\\
\chi_\mathcal{T} &= \frac{\langle \mathcal{T}^2 \rangle - \langle \mathcal{T} \rangle^2}{NT}.
\end{split}
\end{equation}
\noindent
The angular brackets denote the Monte-Carlo average of the quantity. Likewise the average energy per site is given by $ \langle E \rangle = \frac{1}{N} \langle H_{eff} \rangle $, and $\mathcal{T} = \sum_i \mathcal{T}_i$ denotes the discretized average skyrmion density where,
\begin{equation}
 \centering
 \mathcal{T}_{i}  =  \frac{1}{8\pi} [ {\bf S}_i \cdot ({\bf S}_{i+x} \times {\bf S}_{i+y} ) + {\bf S}_i \cdot ({\bf S}_{i-x} \times {\bf S}_{i-y})],
\end{equation}
\noindent
is the local skyrmion density \cite{zhang2016scirep}. The spin structure factor (SSF) is another very useful quantity that helps in identifying the nature of magnetic order. The peak location of the SSF contains the information about the ordering wave vector. Furthermore, it is of direct experimental relevance as the SSF itself is measured in neutron scattering experiments. We compute the component resolved spin structure factor to characterize different ordered magnetic phases. The SSF is given by,

\begin{eqnarray}
 S_f({\bf q}) &=& S^{x}_f({\bf q}) + S^{y}_f({\bf q}) + S^{z}_f({\bf q}), \nonumber \\ \nonumber\\
 S^{\mu}_{f}({\bf q}) &=& \frac{1}{N^2} \bigg \langle \sum_{ij} S^{\mu}_i S^{\mu}_j~ e^{-{\rm i}{\bf q} \cdot ({\bf r}_i - {\bf r}_j)} \bigg \rangle.
\label{eq:ssf}
\end{eqnarray}
\noindent
with $\mu = x,y,z$.

\section{Determination of phase boundaries}
\begin{figure}[h!]
\centering
\includegraphics[width=0.7 \columnwidth,angle=0,clip=true]{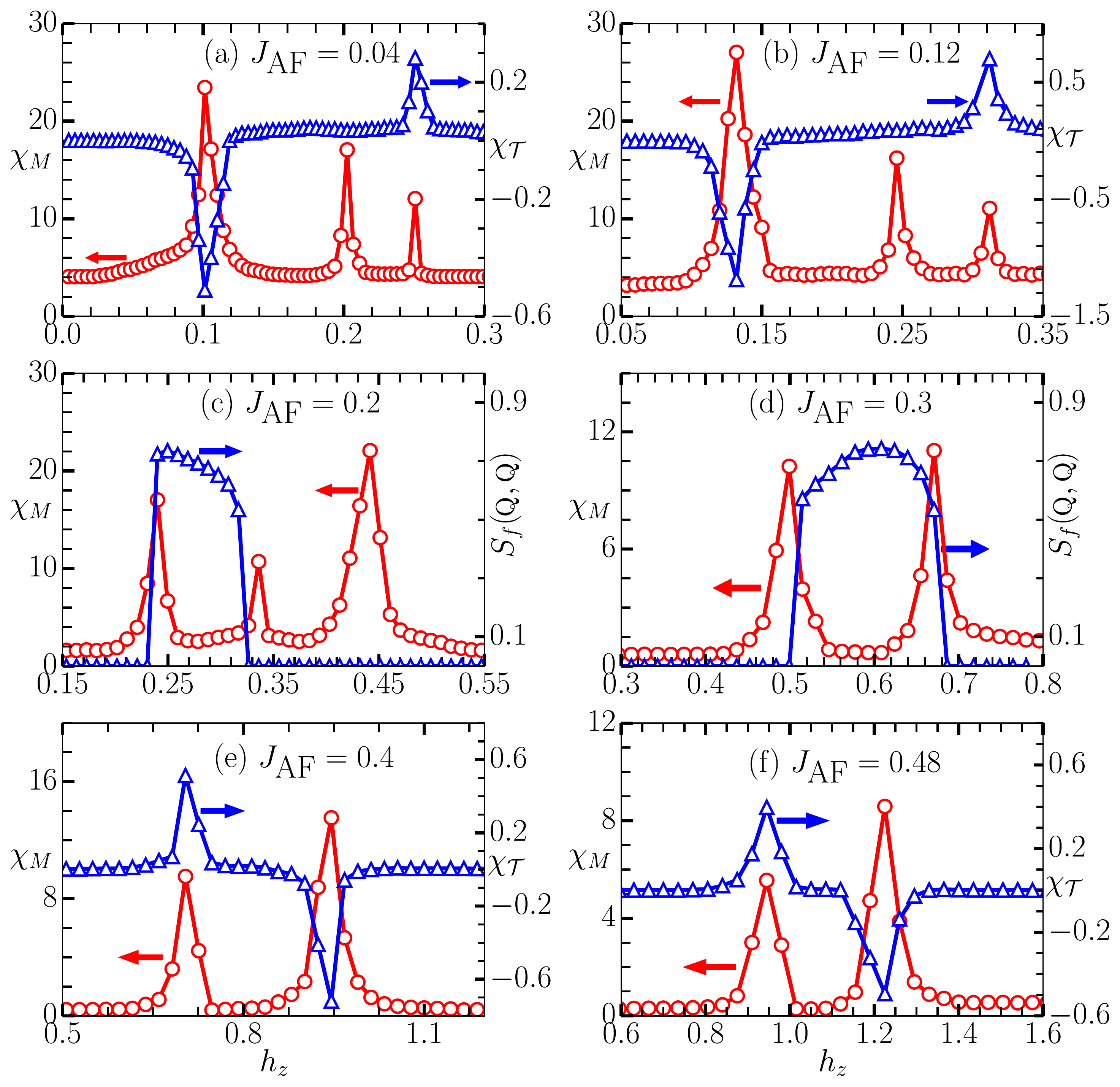}
\caption[Magnetic field dependence of various {\color{red} physical quantities} for different $J_{\text{AF}}$ values.]{Magnetic field dependence of various physical quantities for different $J_{\text{AF}}$ values. (a),(b),(e),(f) shows the variation of magnetic susceptibility ($\chi_M$) [left axis, red circles] and topological susceptibility ($\chi_{\mathcal{T}}$) [right axis, blue triangles] at $J_{\text{AF}} = 0.04, 0.12, 0.4, 0.48$. In (c),(d) in addition to the $\chi_M$ [left axis, red circles], we observe similar order parameter-like behaviour of the SSF peak ($S_f$) strength within pSk state [right axis, blue triangles] at specific ${\bf q}$ locations.
}
\label{sm_fig1}
\end{figure}

The phase boundaries in the phase diagram (Fig. \ref{fig:pd} in the manuscript) are drawn on the basis of variations of magnetic susceptibility ($\chi_M$), topological susceptibility ($\chi_{\mathcal{T}}$) and spin structure factor ($S_f({\bf q})$) as function of $h_z$. At least one of these quantities display a clear anomaly at the phase boundaries. We present the data for a few representative points in the parameter space. In Fig. \ref{sm_fig1} (a), at $J_{\text{AF}}=0.04$ there are clear peaks in $\chi_M$ at $h_z=0.1$, $0.2$, and at $0.25$ demonstrating phase transitions from fDW to FM via pASk and sASk. The $\chi_M$ shows three peaks at that parameter point whereas there are two peaks visible in the $\chi_{\mathcal{T}}$; this is because of the fact that the $\chi_{\mathcal{T}}$ only accounts for the topological
to nontopological transitions and vice versa, namely fDW to pASk and sASk to FM. At $J_{\text{AF}}=0.12$, the $\chi_M$ and $\chi_{\mathcal{T}}$ show similar peaks to quantify the SQ to pASk transition at $h_z=0.13$, and pASk to sASk transition at $h_z \approx 0.25$ (see Fig. \ref{sm_fig1} (b)). At moderate $J_{\text{AF}}$, the ASk-SL state with smallest possible skyrmion size is observered. This 2Q state shows four clear peaks in spin structure factor. We have tracked the intensity of the structure factor ($S_f$) peaks over $h_z$ range for various $J_{\text{AF}}$ values. At $J_{\text{AF}}=0.2, 0.3$ we observe three peaks, corresponds SQ to FM transition via ASk-SL and sASk, in $\chi_M$ along with sudden rise and fall of $S_f$ within pASk state (Fig. \ref{sm_fig1} (c), (d)). At large $J_{\text{AF}}$, we find an exotic antiferromagnetic skyrmion crystal (AF-ASkX) state, arranged on a square lattice, is stable for wide range of $J_{\text{AF}} = 0.34-0.5$. The phase transition from SQ to AF-ASkX is confirmed by peaks in $\chi_M$ and $\chi_{\mathcal{T}}$ (see Fig. \ref{sm_fig1} (e), (f)).

\section{Real-space view of low-temperature magnetic phases}

\begin{figure}
\centering
\includegraphics[width=0.6 \columnwidth,angle=0,clip=true]{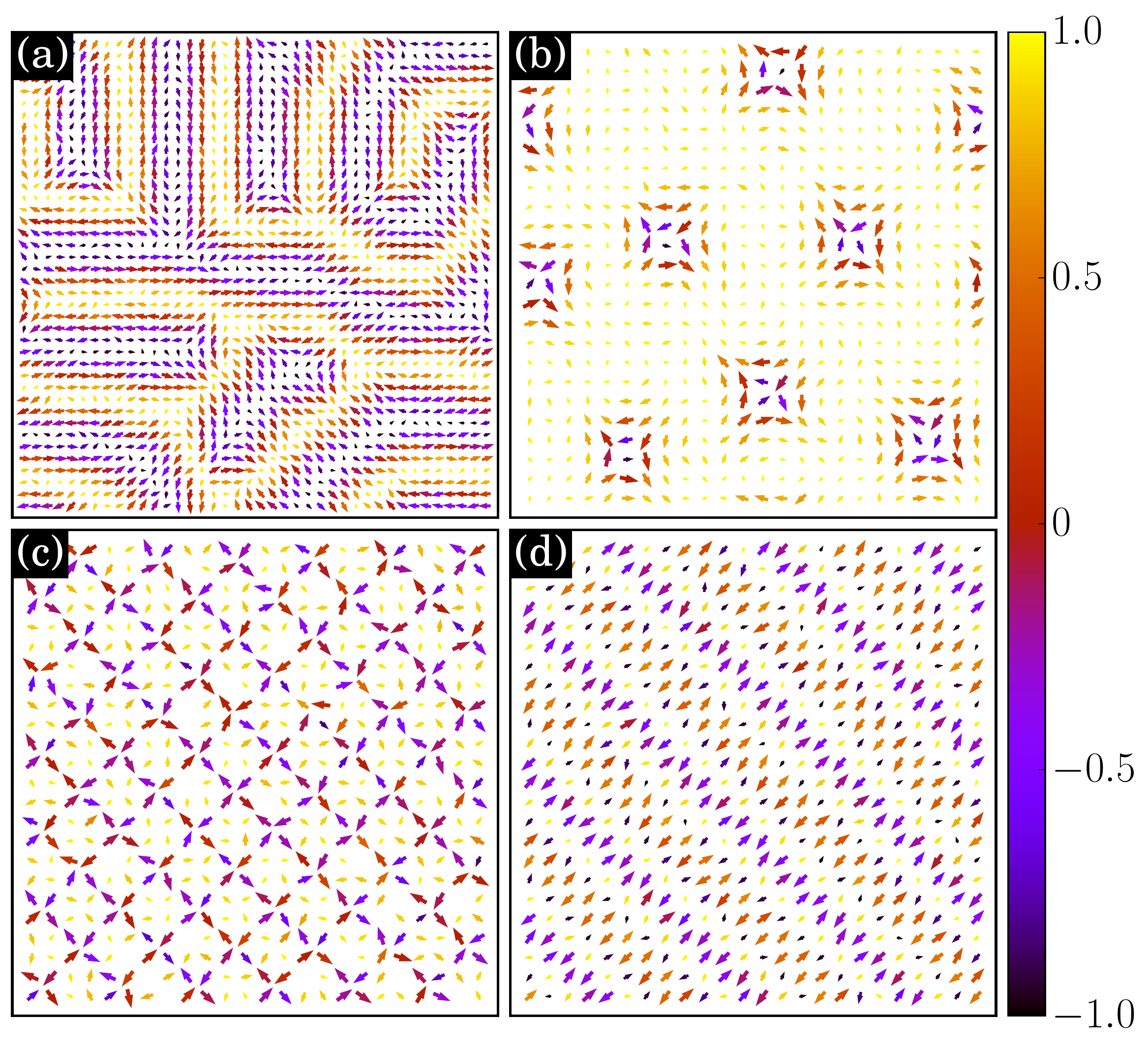}
\caption[Snapshots of various spin configurations obtained in Dresselhaus double-exchange metal in the presence of antiferromagnetic exchange interaction.]{Snapshots of spin configurations obtained at low temperature for, (a) fDW, (b) sASk, (c) ASk-SL, and (d) SQ. The planar components (x, y) of the spin vectors are represented by the arrows while the z component is color coded. For (a) only we show $40 \times 40$ lattice points. For rest of the states we only display $24 \times 24$ section of the original lattice.}
\label{fig:spin_config}
\end{figure}

Typical low temperature real spin configurations for the emergent magnetic states are shown in Fig. \ref{fig:spin_config}. The filamentary domain wall (fDW) state comprises of a domain-wall like structure and the plane of spiralling can bend without any energy cost (see Fig. \ref{fig:spin_config} (a)). The stability of the domain structure is related to an uncommon degeneracy of spiral states that arises from the presence of mutually orthogonal directions of the two DM vectors in our spin model \cite{kathyat2020a}. The Fig. \ref{fig:spin_config} (b) shows a typical sparse antiskyrmion (sASk) state, where the number of skyrmions are reduced with increasing magnetic field. The antiskyrmion square lattice (ASk-SL) states comprises of smallest possible antiskyrmions arranged on a tilted square lattice (see Fig. \ref{fig:spin_config} (c)). For most of the $J_{\text{AF}}$ range we get single-Q (SQ) spiral state as low temperature state at low magnetic field, notice that the spirals are not axial but diagonal (see Fig. \ref{fig:spin_config} (d)).

\section{Effect of $J_{\text{AF}}$}
\begin{figure}
\centering
\includegraphics[width=0.9 \columnwidth,angle=0,clip=true]{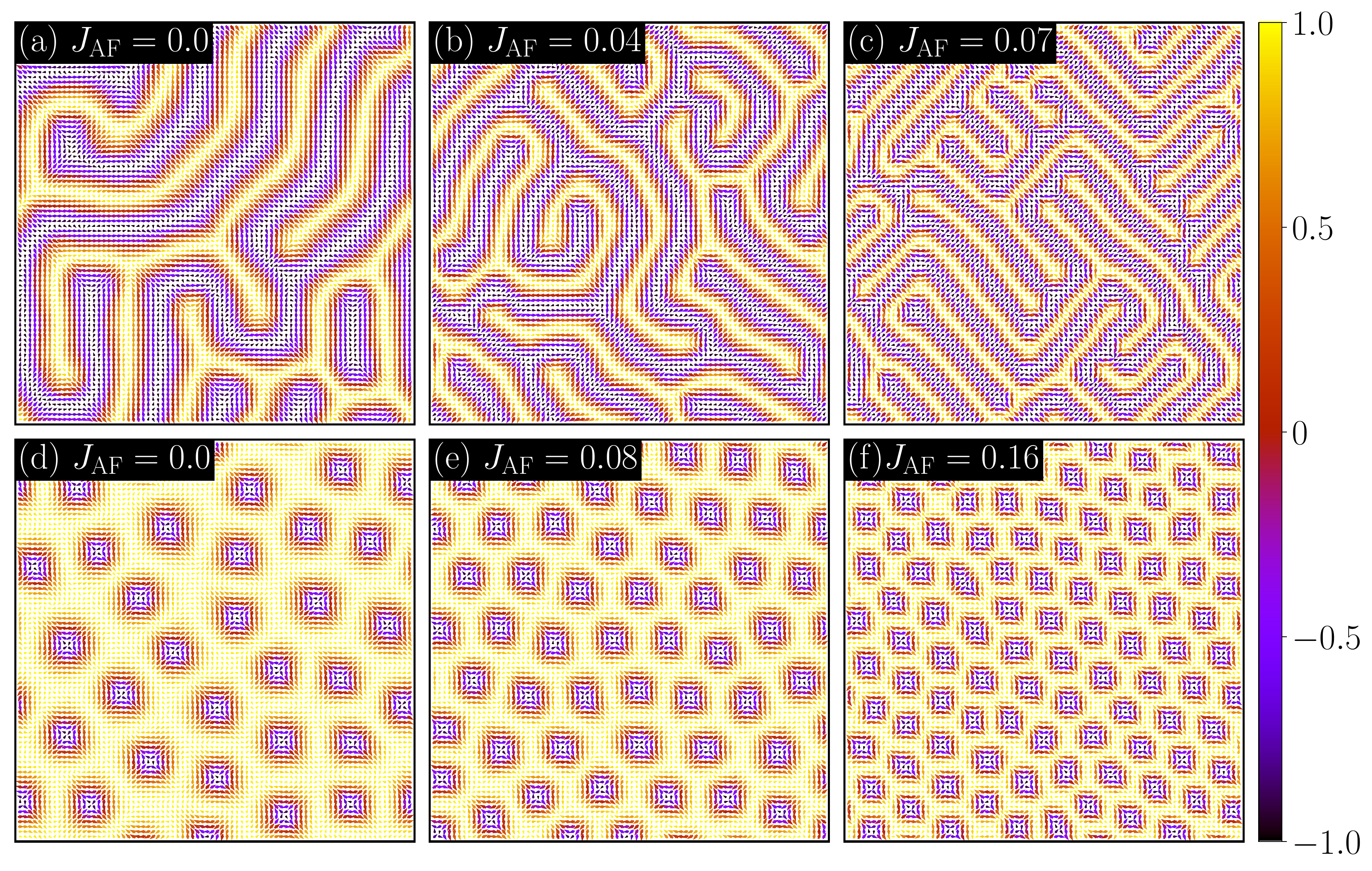}
\caption{Spin configurations of fDW state (a)-(c) and pASk state (d)-(f) for various values of $J_{\text{AF}}$. The planar components ($S_x$, $S_y$) of the spin vectors are represented by the arrows while the $S_z$ is color coded. We display the typical spin configurations obtained at low temperatures in simulations of $H_{\text{eff}}$ (Eq. \ref{eq:ESH_DDE} in the manuscript) on $80 \times 80$ lattice size at $\lambda=0.3$.}
\label{fig:effect_JAF}
\end{figure}

In this section we discuss the effect of $J_{\text{AF}}$ on the magnetic states, fDW and pASk. The results below clearly demonstrate that increasing $J_{\text{AF}}$ has effects similar to those known for increasing $\lambda$. We can observe from Fig. \ref{fig:effect_JAF} (a)-(b), the width of the filamentary domain walls get reduced with increasing value of $J_{\text{AF}}$ at $h_z = 0.0$, the similar effect has been reported in our previous work where increasing spin-orbit coupling leads to reduction in width of domain walls \cite{kathyat2020a}. A crucial difference here seems to be a preference for diagonally oriented domain walls for larger values of $J_{\text{AF}}$. This is understandable as the AFM coupling, by itself, prefers a diagonal $(\pi,\pi)$ ordering. Further, in Fig. \ref{fig:effect_JAF} (d)-(f), we observe that as we increase the antiferromagnetic exchange coupling $J_{\text{AF}}$ at $h_z=0.12$ the size of the emerging antiskyrmion reduces. This effect is also similar to that known for the dependence of skyrmion size on the strength of spin-orbit coupling.

\pagebreak
\twocolumngrid
\bibliographystyle{apsrev4-2} 
%

\end{document}